\documentclass[twocolumn,pra,nofootinbib,superscriptaddress,aps,floatfix]{revtex4-1}
\usepackage{graphicx,epsfig}
\usepackage{bm}
\usepackage{amsmath}
\usepackage{amssymb}

\def\la{\langle}
\def\ra{\rangle}
\newcommand{\tr}[0]{\ensuremath{\mathrm{Tr}}}
\newcommand{\trf}[0]{\ensuremath{\mathrm{Tr_f}}}
\newcommand{\ket}[1]{\ensuremath{\left|{#1}\right\rangle}}
\newcommand{\bra}[1]{\ensuremath{\left\langle{#1}\right|}}
\newcommand{\braket}[1]{\ensuremath{\left\langle{#1}\right\rangle}}

\begin{document}
\title{\bf{Zeno physics in ultrastrong circuit QED}}

\author{I. Lizuain}
\affiliation{Departamento de Qu\'{\i}mica F\'{\i}sica, Universidad del Pa\'{\i}s Vasco - Euskal Herriko Unibertsitatea, Apdo. 644, 48080 Bilbao, Spain}

\author{J. Casanova}
\affiliation{Departamento de Qu\'{\i}mica F\'{\i}sica, Universidad del Pa\'{\i}s Vasco - Euskal Herriko Unibertsitatea, Apdo. 644, 48080 Bilbao, Spain}

\author{J. J. Garc\'{\i}a-Ripoll}
\affiliation{Instituto de F\'{\i}sica Fundamental, CSIC, Serrano 113-bis, 28006 Madrid, Spain}

\author{J. G. Muga}
\affiliation{Departamento de Qu\'{\i}mica F\'{\i}sica, Universidad del Pa\'{\i}s Vasco - Euskal Herriko Unibertsitatea, Apdo. 644, 48080 Bilbao, Spain}

\author{E. Solano}
\affiliation{Departamento de Qu\'{\i}mica F\'{\i}sica, Universidad del Pa\'{\i}s Vasco - Euskal Herriko Unibertsitatea, Apdo. 644, 48080 Bilbao, Spain}
\affiliation{IKERBASQUE, Basque Foundation for Science, 48011 Bilbao, Spain}

\date{\today}


\begin{abstract}
We study the Zeno and anti-Zeno effects in a superconducting qubit interacting strongly and ultrastrongly with a microwave resonator. Using a model of a frequently measured two-level system interacting with a quantized mode, we predict different behaviors and total control of the Zeno times depending on whether the rotating-wave approximation can be applied in the Jaynes-Cummings model, or not. We exemplify showing the dependence of our results with the properties of the initial field states.
\end{abstract}

\maketitle

\section{Introduction}

Frequent measurements of a quantum mechanical system may cause its evolution to change, slow down or even freeze. This phenomenon, known as the quantum Zeno effect~\cite{MS77,pascazio08review}, implies in particular that an atom that is being periodically or continuously monitored cannot decay, or will do it at a slower rate. This suppression of spontaneous emission was first observed in trapped ion experiments~\cite{itano_zeno}, followed by observations in the suppression of Rabi oscillations in cavity QED~\cite{BDSKDBRH08}, the decay of ultracold atoms~\cite{ket06}, and has been studied associated to photodetection in circuit QED~\cite{helmer09}.

Far from being just an interesting consequence of quantum mechanics, Zeno physics is also a useful tool in quantum control: it can be used to protect quantum information in certain subspaces~\cite{barenco97}, to suppress decoherence during quantum gates~\cite{beige00}, or even as a means of doing efficient quantum search~\cite{childs02}. Interestingly, it has also been shown that frequent measurements can produce the opposite effect~\cite{KK00nature,scully08nature}, enhancing the decay of a quantum system, in what is known as anti-Zeno effect. This other possibility has potential applications for cooling~\cite{kurizki08nature} and has been experimentally demonstrated~\cite{raizen_AZE}.

Superconducting circuits are good candidates for studying Zeno physics~\cite{gambetta08}. The equivalent of an atom interacting with the electromagnetic field in free space or in a cavity~\cite{itano_zeno} is replaced by a superconducting qubit interacting with an open or closed microwave transmission line~\cite{walraff04}. By monitoring the state of the qubit, or of the cavity, one expects to have the possibility of freezing or accelerating the evolution. To our advantage, the time scales involved in the circuit QED dynamics and its measurements are now much larger as compared to the optical domain, and can be resolved with ordinary electronics. Furthermore, all energy scales of related experiments, including couplings, qubit and resonator frequencies, can be engineered and tuned at will. This opens the possibility of reaching the ultrastrong coupling regime~\cite{bourassa09,abdumalikov08,niemczyk10,forn-diaz10}, where the rotating-wave approximation (RWA) breaks down and the system cannot be described by a Jaynes-Cummings dynamics~\cite{jc63}.

The goal of this work is indeed to study the Zeno physics in the qubit-cavity coupling of circuit QED with and without the RWA, that is, in the strong and ultrastrong coupling regimes. We will do it by using the model of a two-level system of frequency $\omega_0$ interacting with a single-mode harmonic oscillator of frequency $\omega,$ and assuming periodic measurements of the qubit with intervals $\delta t.$ As we will show, there are two possible regimes. First, the case where the measurement is faster than the vacuum Rabi coupling $g$ of the combined qubit-resonator system, but slower than the qubit and resonator frequencies: $g \ll (\delta t)^{-1} \ll \{ \omega,\omega_0 \}$. Here, the RWA is valid and we recover the usual slow down of the qubit decay. Second, the case where the qubit is measured more rapidly than any existing energy scale, $(\delta t)^{-1} \gg \{g, \omega,\omega_0 \}$. Here, the RWA breaks down and we can observe both an enhancement of the Zeno or even an anti-Zeno effect, in which the decay of the excited state is accelerated with respect to the RWA case.

The structure of the paper is as follows. We introduce the model of a qubit interacting with a single-mode resonator and derive the evolution of the system under repeated measurements. It will be shown that fast enough measurements induce an evolution of the qubit population that is well approximated by an exponential decay, a trace of the Zeno physics. The rate of the effective decay is shown to be different for a model that follows the RWA and for one that does not. We will discuss the conditions for these differences and explore their consequences using coherent and squeezed states, both analytically and numerically. Finally, we will suggest realistic implementations of the proposed ideas in current setups of quantum circuit technologies.

\section{Survival probability and Zeno Effect}

Let us consider the dynamics of a two level system, $\ket{\mathrm{g}}$ and $\ket{\mathrm{e}},$ separated by a gap $\omega_0,$ and coupled to a harmonic oscillator of frequency $\omega.$ The whole system models a superconducting circuit coupled to a high-Q microwave resonator, and corresponds to the Hamiltonian ($\hbar = 1$)
\begin{eqnarray}
\label{gral_hamiltonian}
H&=&g_{\mathrm{r}}\left(\sigma_+a+\sigma_-a^\dag\right)+g_{\mathrm{b}} \left( \sigma_+ a^\dag + \sigma_- a \right) \\
&+&\omega a^\dag a+\frac{\omega_0}{2}\sigma_z . \nonumber
\end{eqnarray}

The couplings associated with the co-rotating Jaynes-Cummings (denote by ``r'', from red) and counter-rotating Anti-Jaynes-Cummings (denote by ``b'', from blue) are equal, $g=g_{\mathrm{r}}=g_{\mathrm{b}}$, but we express them separately to trace their physical consequences. It is known that when $\omega$ and $\omega_0$ are much larger than the coupling strength,  $g$, the RWA applies and formally $g_{\mathrm{b}}=0,$ yielding the Jaynes-Cummings model~\cite{jc63}.

We will focus on the {\it survival probability}, $P_{\mathrm{e}}(t)$: the probability of preparing the two-level system in a given state, say $\ket{\mathrm{e}}$, and finding it  unaltered at a given time $t$. Consider a system described by a qubit in an excited state and the cavity in an arbitrary state $\rho_f,$ that is $\rho(0)=\ket{\mathrm{e}}\bra{e}\otimes\rho_f$. The survival probability of this excited state at the first measurement is a function of time
\begin{eqnarray}
P_{\mathrm{e}}(t)&=&\tr\left[\rho(t)|\mathrm{e}\ra\la \mathrm{e}|\right]\\
&=&\tr\left[e^{-iH t}\rho(0)e^{iH t}|\mathrm{e}\ra\la \mathrm{e}|\right]\nonumber\\
&=&\tr\left[\sum_{n=0}^\infty\frac{(-i t)^n}{n!}\left[H,\rho(0)\right]_n|\mathrm{e}\ra\la \mathrm{e}|\right] ,
\nonumber
\end{eqnarray}
where $[H,\rho(0)]_n:=[H,[H,\hdots[H,\rho(0)]]]$ comes from the Baker-Campbell-Hausdorff formula.

The short-time dynamics can be determined by the lowest order non-vanishing contribution in the previous sum. It can be shown that the linear term, ${\cal O}(t),$ vanishes for \emph{arbitrary} initial atomic states. Therefore the survival probability for short times decays at most quadratically
\begin{equation}
\label{quadratic_decay}
P_{\mathrm{e}} (t) \approx 1-\left(\frac{t}{\tau_Z}\right)^2,
\end{equation}
with a scale $\tau_Z$ known as the  ``Zeno time''~\cite{SRM94}. If instead of performing just one measurement, one asks for the {\it repeated survival probability}, $P_{\mathrm{e},N}(N\delta t)$, where $N$ measurements at regular intervals $\delta t$ were made, the result is no longer a quadratic but rather the exponential decay~\cite{facchi01}
\begin{equation}
P_{\mathrm{e},N}(t=N\delta t) = \left[1-\left(\frac{\delta t}{\tau_Z}\right)^2\right]^N\approx e^{-\gamma_{\mathrm{eff}} t} .
\label{exponential}
\end{equation}

The effective decay rate, $\gamma_{\mathrm{eff}},$ and resulting lifetime, $\tau_{\mathrm{eff}}$, depend on the Zeno time
\begin{equation}
\label{effective_decay_zeno}
\gamma_{\mathrm{eff}}=1/\tau_{\mathrm{eff}}=\delta t /\tau_Z^2,
\end{equation}
and for frequent enough measurements, we will observe an effective freezing of the dynamics of the measured system, the proper Quantum Zeno effect~\cite{MS77}.

The previous general considerations can be particularized for the Hamiltonian (\ref{gral_hamiltonian}) to obtain the Zeno time
\begin{eqnarray}
\label{zeno_time_general}
\tau_Z^{-2}
&=&g_{\mathrm{r}}^2\braket{aa^\dag}_{\rho_f} +g_{\mathrm{b}}^2\braket{a^\dag a}_{\rho_f}\\
&+&g_{\mathrm{r}}g_{\mathrm{b}}\braket{a^2+a^{\dag 2}}_{\rho_f}\nonumber,
\end{eqnarray}
where the expectation values $\braket{X}_{\rho_f} = \trf\left[\rho_fX\right]$ are taken over the initial state $\rho_f.$ Let us note that the Zeno time and the resulting decay rate are not only very sensitive to the initial field states, $\rho_f,$ but also to the precise form of the Hamiltonian. If the RWA is valid,  $g_{\mathrm{b}}=0,$ we obtain that the time depends on the total number of photons, $\tau_Zº^{-2} \propto \braket{a^\dag a}+1,$ while in the ultrastrong coupling regime it also involves other field quadrature momenta, $\braket{(a+a^\dag)^2}.$

The previous derivations are valid only under certain restrictions. In particular, the interval between measurements has to be short, so that we can replace the polynomial expression in (\ref{exponential}) with an exponential. This can be done if the time $\delta t$ is much shorter than the relevant periods. In the case of a strong coupling regime with RWA, we will only need that this is shorter than the Rabi frequency, that is about $\delta t \ll 1/g.$ However, for ultrastrong couplings in which the RWA does not apply, we must impose faster measurements $\delta t \ll \{ 1/\omega , 1/g \}.$

\begin{figure}[t]
\begin{center}
\includegraphics[width=0.9\linewidth]{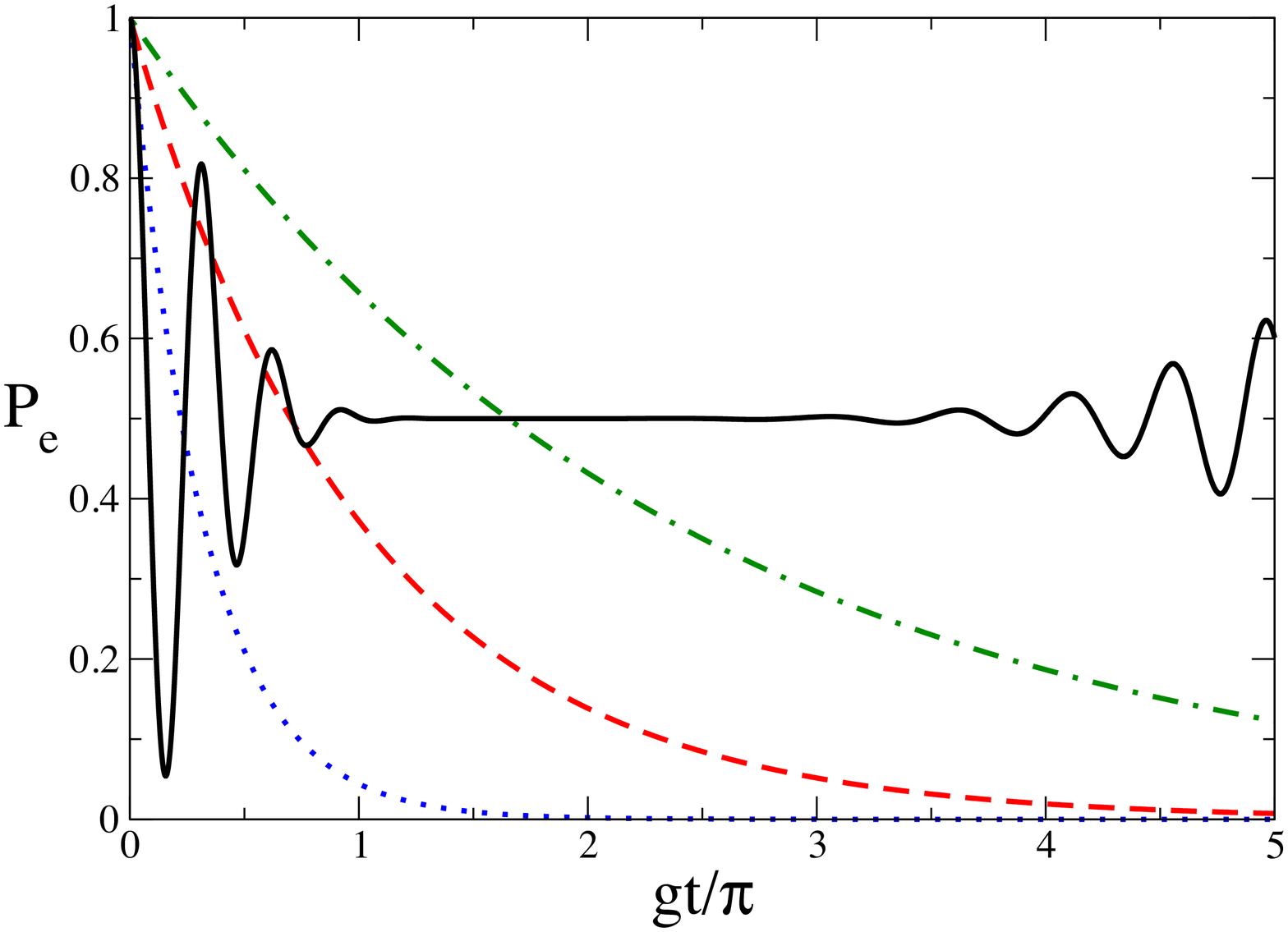}
\caption[]{(Color online) Zeno effect in the strong coupling regime, $g/\omega=0.05,$ for an initial state with an excited qubit and the resonator in a coherent state, $\ket{\mathrm{e}}\otimes\ket{\alpha},$ with $|\alpha|^2=9$ photons on average. In solid line we plot the collapses and revivals under free evolution in the RWA. The remaining lines correspond to the survival probability of the excited state under periodic measurements of the qubit, repeated at intervals $\delta{t}=0.01\pi g^{-1}$. 
If the RWA is applied, the usual Zeno time is obtained (dashed, red). If the RWA is not applied, we can either slow down the decay, setting the phase of the coherent state $\theta=\pi/2$ (green, dash-dot) or increase it $\theta=0$ (dotted, blue). For sufficiently fast measurements, these numerical results and those obtained from the analytical expressions (\ref{coherent_state_effective_decay_rate}) and (\ref{coherent_state_effective_decay_rate_RWA}) are practically the same.}
\label{coherent_state_fig}
\end{center}
\end{figure}

It is also important to note that the Zeno time (\ref{zeno_time_general}), and thus the effective decay rate (\ref{effective_decay_zeno}),
depends only on the initial field state $\rho_f,$ which is assumed constant. We may confirm this by performing a similar calculation as before. After $N$ observations of the internal state of the two-level system every $\delta t$, the field state will be
\begin{eqnarray}
\rho_f^{(N)}&\approx&\rho_f^{(N-1)}+\frac{(-i\delta t)^2}{2!}
\tr_s\left[H,\left[H,\rho_f^{(N-1)}\otimes|\mathrm{e}\ra\la \mathrm{e}|\right]\right]\nonumber
\end{eqnarray}
where $\tr_s$  denotes the trace over the two-level system. By measuring the internal state of the two-level system and restricting ourselves to the survival probability, the field state is also frozen, i.e., the initial field state after each measurement is the initial field state plus some higher order corrections, as confirmed by the numerics.

\section{Different initial states}

Let us now study the changes in the Zeno effect depending on the initial resonator state, as well as the extent of the differences between RWA and non-RWA models. We begin with the Fock state, which has a well defined number of photons,  $\ket{n}$ and gives an effective decay rate
\begin{eqnarray}
\left(\gamma_{\mathrm{eff}}\right)_n&=&
\delta t\left[g_{\mathrm{r}}^2(n+1)+g_{\mathrm{b}}^2n\right]
\end{eqnarray}
that gives a difference of about a factor of two between the RWA and the non-RWA results.

Much more interesting is the case when the field is assumed to be in a coherent state $\ket{\alpha}$ \cite{Mun1,Mun2}, with mean photon number $\bar{n}$ and a phase $\theta,$ that is $\alpha=\bar{n}^{1/2}e^{i\theta}.$ The effective decay rate becomes
\begin{eqnarray}
  \left(\gamma_{\mathrm{eff}}\right)_\alpha&=&\delta t\left[g_{\mathrm{r}}^2(1+|\alpha|^2)+g_{\mathrm{b}}^2|\alpha|^2\right.\nonumber\\
  &+&\left.2g_{\mathrm{r}}g_{\mathrm{b}}|\alpha|^2\cos{2\theta}\right].
\end{eqnarray}
Out of the two experimentally accessible regimes we then have that
\begin{eqnarray}
\label{coherent_state_effective_decay_rate}
\left(\gamma_{\mathrm{eff}}\right)_\alpha&=&\delta t g^2\left(1+4|\alpha|^2\cos^2\theta\right),\\
\left(\gamma_{\mathrm{eff}}^{\mathrm{RWA}}\right)_\alpha&=&\delta t g^2\left(1+|\alpha|^2\right).
\label{coherent_state_effective_decay_rate_RWA}
\end{eqnarray}
Note that only when the RWA is not applied, the Zeno effect becomes sensitive to the phase of the coherent state. 
\begin{figure}[t]
\begin{center}
\includegraphics[width=0.9\linewidth]{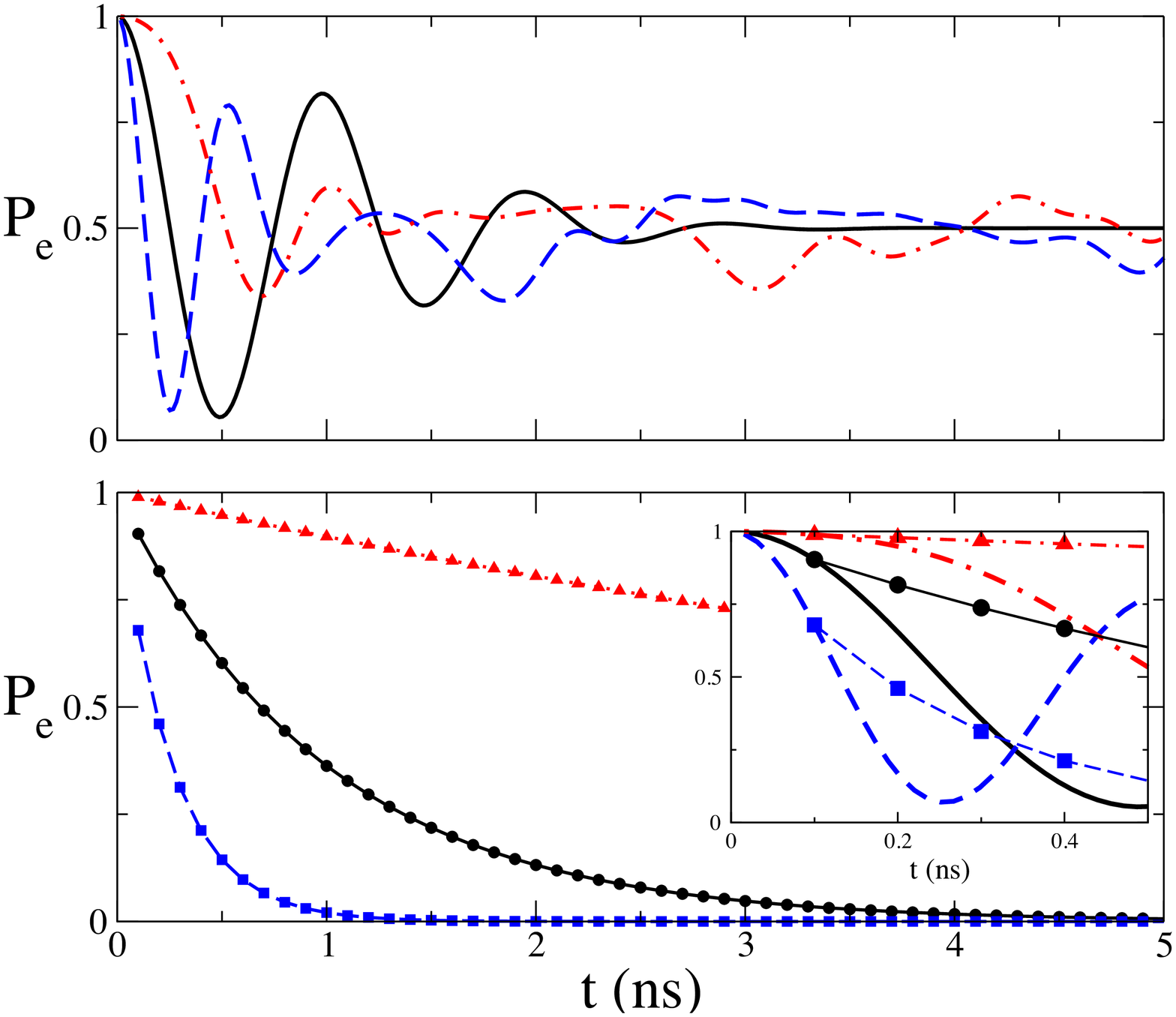}
\caption[]{(Color online) Zeno effect in the ultrastrong coupling regime, when $g$, $\omega$, $\omega_0=1$ GHz. (top) Free evolution of the excited state population in the qubit and (bottom) survival probability of the excited state for periodic measurements with $\delta{t}=0.1\omega^{-1}=0.1$ns (each symbol represents a measurement). (inset) Comparison between both plots. The solid line corresponds to a RWA model (black, solid, circles), while the other lines start from a coherent state with 9 photons and either $\theta=0$ (blue, dashed, squares) or $\theta=\pi/2$ (red, dash-dot, triangles).}

\label{ultrastrong_fig}
\end{center}
\end{figure}

Finally, we will consider a field initially in a vacuum squeezed state, $\ket{\psi_\xi},$ that is
\begin{equation}
\label{sq_operator}
\ket{\psi_\xi}=e^{\frac{1}{2}(\xi^* a^2-\xi a^{\dag 2})}\ket{0},
\end{equation}
where $\xi=re^{i\theta}$ is the squeezing parameter. The decay rates, beyond and within the RWA, now become
\begin{eqnarray}
\label{decay_rates_vacuum_squeezed_no_RWA}
\left(\gamma_{\mathrm{eff}}\right)_\xi&=&\delta t \ g^2\left(\cosh{2r}-\cos\theta\sinh{2r}\right)\\
\left(\gamma_{\mathrm{eff}}^{\mathrm{RWA}}\right)_\xi&=&\delta t\ g^2(\cosh r)^2
\label{decay_rates_vacuum_squeezed_RWA} .
\end{eqnarray}

We have found that for coherent and squeezed states, the decay rate depends on the field phase. This dependence is enhanced in the non-RWA regime, which allows us for a total control of the decay rate. In the case of coherent states, we can make the decay rate independent of the number of photons. In the case of squeezed states we can change the way the decay rate depends on the squeezing, $r$: for the RWA or for $\theta=\pi/2$ the Zeno decay increases as $1+r^2$ while for other values of $\theta$ we may achieve $\gamma_{\mathrm{eff}}=0$ or make it decrease with the squeezing parameter as $1-2r$ if $\theta=0$.
We may regard the later as an example of the quantum Zeno effect and the former as an example of the so-called anti-Zeno effect, where the dynamics of the system is accelerated rather than slowed down by frequent measurements. Note that this is an acceleration with respect to the RWA result, not to the natural decay rate since the system is not unstable. The transition from Zeno to anti-Zeno regimes in truly unstable systems is discussed in Ref.~\cite{Modi2007}. A possible definition of the Zeno and anti-Zeno time scales in oscillatory systems (not unstable) is given in Ref.~\cite{facchi01}.

The previous results can be numerically verified both in the strong and ultrastrong coupling regimes. In Figs.~\ref{coherent_state_fig} we plot the evolution of a qubit-resonator system in which the coupling strength is small enough to satisfy the RWA in the free-evolution case, $g=0.05 \omega$ with $\omega=\omega_0.$ Beginning with a coherent state in the field and an excited state in the qubit, the results are collapses and revivals of the excited state population, a phenomenon that can be described with the Hamiltonian~(\ref{gral_hamiltonian}) with $g_{\mathrm{b}}=0.$ However, if we perform measurements with a periodicity $\delta{t}=0.2\pi\omega^{-1}=0.01\pi g^{-1},$ we observe that the survival probability of the excited state follows an exponential law (\ref{exponential}) with a decay rate that is not the one given by the RWA (red, dashed line), but rather the one in Eq.~(\ref{coherent_state_effective_decay_rate}). In particular, the decay rate can be enhanced by setting the resonator in a coherent state with $\theta=0,$ or decreased, by choosing $\theta=\pi/2.$ A similar result is observed for ultrastrong couplings, $g=\omega=\omega_0,$ as in Fig.~\ref{ultrastrong_fig}.

\begin{figure}
  \centering
  \includegraphics[width=0.9\linewidth]{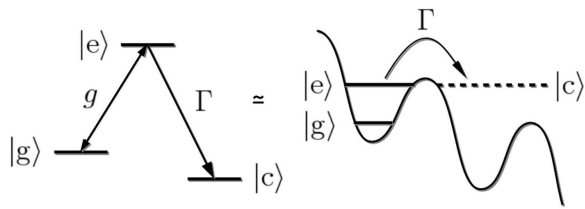}
  \caption{A phase qubit can be used to explore the Zeno effect: the escape rate of the unstable level $\ket{\mathrm{e}}$ is equivalent to a continuous measurement of that state.}
  \label{fig:phase-qubit}
\end{figure}

\section{Experimental implementation}

The main question that arises when looking for an experimental implementation of the previous ideas is how to achieve the regime of fast repeated measurements in superconducting circuits. As discussed before, we need to probe both a regime in which the time interval $\delta t$ is much smaller than the vacuum Rabi period but the RWA remains valid, and a regime in which the measurements take place much faster than any other time scale of our system. Given that current measurement devices, as SQUIDs, need nanoseconds to operate, we would have to relax all energy scales, $\omega$, $\omega_0$, and $g$, and risk being affected by thermal fluctuations. However, this is not unconceivable because we could design an experiment near the largest frequency $\omega=2\pi\times 200$ MHz, which gives a vacuum Rabi period of about 5 ns. The cavity will then be populated by an average of 2 thermal photons, but the features of the Zeno effect should remain visible.

Another possibility is to work in a framework of continuous measurements, like in fluorescence experiments with trapped ions~\cite{itano_zeno}. In that case, we suggest the use of a resonator interacting with a phase qubit~\cite{martinis85} whose excited state, $\ket{\mathrm{e}}$ may decay into the continuum at a rate $\Gamma$, [See Fig.~\ref{fig:phase-qubit}]. By studying the repeated survival probability of the qubit \textit{ground state}, we should observe also a Zeno effect, where the incoherent decay plays the role of a continuous measurement. According to Refs.~\cite{schulman,MEAL08}, the decay channel of the qubit excited level gives an effective measurement interval $\delta{t}=4/\Gamma.$ By continuously monitoring whether the phase qubit switches to a voltage state, we will be able to determine the survival probability of the ground state. The phenomenology should be equivalent to the previous cases, with the advantage that now $\omega$ can be larger, of the order of GHz, due to the flexibility in tuning the decay rate $\Gamma.$

Arguably, the most promising setup would consist on a superconducting qubit interacting via a tunable coupling with the microwave resonator~\cite{peropadre09}. By controlling the interaction we should be able not only to choose between strong and ultrastrong couplings, but also to completely deactivate the interaction for brief periods of time in order to measure the system. This switching on and off can be done with sub-nanosecond resolution, allowing us to reach the regime $\delta t \ll \{ 1/\omega,1/\omega_0,1/g \},$ and the whole system should behave just as if we were able to instantaneously monitor the superconducting qubit. 

Summing up, in this work we have shown the difference in the Zeno dynamics between a quantum optical system that obeys the rotating-wave approximation and one that does not. In the RWA case, the Zeno effect only depends on the total number of photons of the electromagnetic field that interacts with the monitored qubit. In the non-RWA case, the Zeno decay may be completely suppressed or enhanced (anti-Zeno), by suitably preparing the phase of a coherent or squeezed field. Both regimes should be observable in quantum circuits consisting on a qubit interacting with a high-quality microwave resonator. Note that in contrast to previous works~\cite{zubairy08}, the choice of the physical system makes the discrimination of RWA/non-RWA more accessible and also avoids the subtleties and problems associated with a continuum of modes.

The authors acknowledge funding from UPV-EHU Grant GIU07/40, Spanish MICINN project FIS2009-12773-C02-01, EuroSQIP and SOLID European projects, ESF network program ``AQDJJ'',  Basque Government Grants BF108.211 and IT472-10, Spanish MEC Project FIS2006-04885 and CSIC Project 200850I044.

\end{document}